\documentclass[apj,twocolumn]{emulateapj_mod}

\makeatother

\newcommand{\lsim}{\ \raise -2.truept\hbox{\rlap{\hbox{$\sim$}}\raise5.truept
        \hbox{$<$}\ }}
\newcommand{\gsim}{\ \raise -2.truept\hbox{\rlap{\hbox{$\sim$}}\raise5.truept
        \hbox{$>$}\ }} 
\newcommand{\kms}{km~s$^{-1}$}
\newcommand{\cm}{cm$^{-2}$}
\newcommand{\grb}{GRB~020813}
\newcommand{\feii}{$\log N_{\rm FeII}$}
\newcommand{\mnii}{$\log N_{\rm MnII}$}

\slugcomment{ApJ in press}
\shorttitle{Dust depletion and extinction in \grb}
\shortauthors{Savaglio \& Fall}

\begin{document} 

\input epsf 

\title{DUST DEPLETION AND EXTINCTION IN A GAMMA-RAY BURST AFTERGLOW}

\author{Sandra Savaglio \altaffilmark{1}}
\affil{Johns Hopkins University, 3400 North Charles Street,
Baltimore, MD21218; savaglio@pha.jhu.edu}

\author{S. Michael Fall}
\affil{Space Telescope Science Institute, 3700 San Martin Drive,
Baltimore, MD21218; fall@stsci.edu}

\altaffiltext{1}{On leave of absence from INAF, Osservatorio Astronomico di
Roma, Italy} 

\begin{abstract} 

We put stringent constraints for the first time on the dust properties
in the circumburst medium of a gamma-ray burst (GRB) afterglow. This
is based on the optical spectrum of \grb\ ($z=1.255$), obtained with
Keck I LRIS 4.65 h after the burst. From the
absorption lines in the spectrum, we derive very high column densities
for six heavy elements with different refractory properties. The
relative abundances resemble the dust depletion patterns in the Milky
Way, from which we infer a visual extinction of $A_V\simeq0.4$ 
and $A_V>0.3$ at
95\% confidence level. However, the high columns of metals and dust
contrast with an observed UV continuum spectrum that is
remarkably close to a power law of the form $F_\nu \propto
\nu^{-0.9}$, with no sign of curvature, or a 2200 \AA\
extinction feature, suggesting low reddening. The Milky Way or
Magellanic Cloud reddenings are possible only for very low
visual extinctions ($A_V<0.08$ or $A_V<0.2$, respectively at 95\%
confidence), inconsistent with the $A_V$ values inferred from the
depletion analysis.  If we assume a GRB intrinsic spectrum and an
extinction law of the forms $F_\lambda^i = F_V
(5500/\lambda)^\alpha$ and $A_\lambda= A_V (5500/\lambda)^\gamma$, we
obtain (for $A_V=0.4$) the constraints from continuum
spectrum: $\gamma<0.85$ and $\alpha<1.72$.

\end{abstract}

\keywords{cosmology: observations -- gamma rays: bursts -- galaxies: 
abundances -- ISM: dust, extinction}

\section{Introduction}

Long-duration gamma-ray bursts (GRBs) are probably associated with
core-collapse supernovae (Hjorth et al. 2003a) and can be used to
probe the interstellar medium of their star-forming environments.  One of
the open issues on the nature of GRBs is the dust properties in the
circumburst medium. On the one hand, high heavy-element column
densities have been measured in a few GRB afterglow spectra (Savaglio
et al. 2003; Vreeswijk et al. 2004), suggesting that the dust
content might be important, especially if star formation is very
active. On the other hand, the spectral energy distribution (SED) of
other GRB afterglows indicates that dust reddening is surprisingly low
(Galama \& Wijers 2001; Hjorth et al. 2003b). Up to now, a
simultaneous analysis of heavy-element abundances to measure the dust
depletion, and of a suitable UV/optical spectrum to estimate the dust
extinction law, in the same GRB, has never been done. Afterglows fade
so rapidly that obtaining  high-quality spectra, for a full
investigation of the dust properties, is very difficult.

\begin{figure*}\label{f1}
\centerline{{\epsfxsize=15cm \epsfbox{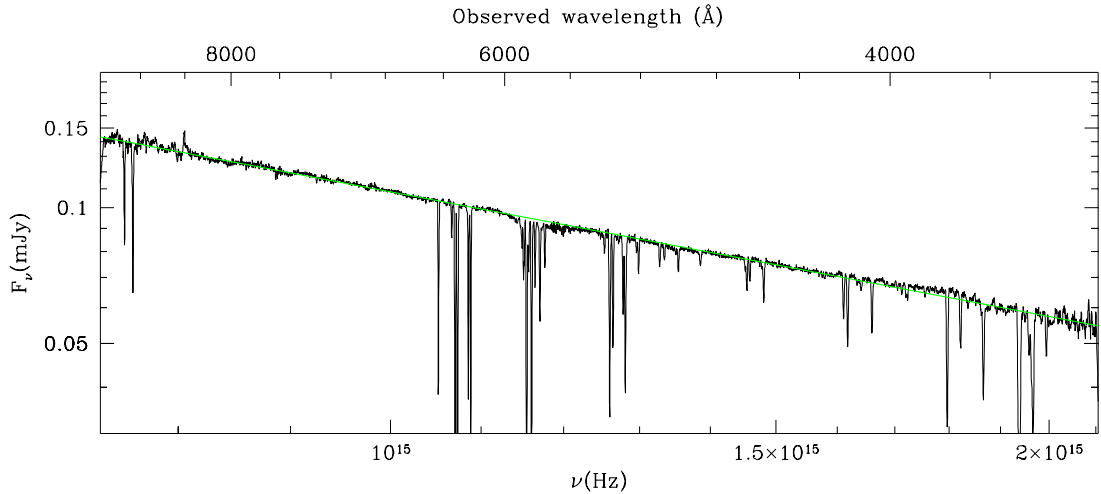}}}
\figcaption[f1]{Optical spectrum ($\lambda\lambda=1420-4080$ \AA\ rest
frame) of \grb\ ($z=1.255$; Barth et al.\ 2003) after correction for
Galactic extinction with $E_{B-V}=0.111$. The straight line is
the power-law best-fit $F_\nu = \nu^{-0.918\pm0.001} \times
10^{12.808\pm0.011}$.}
\end{figure*}

The spectrum of \grb, obtained recently by Barth et al.\ (2003),
offers us this opportunity.  The optical spectrum has a very high 
signal-to-noise ratio (S/N),
is very well flux-calibrated, and shows very little deviation from a
perfect power law over a large wavelength interval ($F_\nu \propto
\nu^{-0.92}$ in $\lambda\lambda=1430-4020$ \AA\ rest frame;
Fig.~1). Moreover, the absorption lines of numerous different
elements, with different refractory properties, have been detected
with exceptional accuracy, thanks to the excellent spectral
resolution.  By visually inspecting the GRB spectrum, it can already
be noticed that there is no sign of the 2200 \AA\ extinction feature
($\nu \approx 1.4\times10^{15}$ Hz), or curvature on even larger
wavelength intervals, suggesting either an extinction law
different from the Galactic type or a low dust content.

Our approach to studying the dust depletion is similar to what has
been done routinely for years for strong absorption systems detected
in QSO spectra, the so-called damped Ly$\alpha$ (DLA) systems. The
spectrum of \grb\ shows similar absorption features, most of which are
associated with low-ionization species, so we consider this absorption
system a GRB-DLA.  Barth et al. (2003) detected seven low-ionization
species (\ion{Fe}{2}, \ion{Si}{2}, \ion{Zn}{2}, \ion{Cr}{2},
\ion{Mn}{2}, \ion{Mg}{2} and \ion{Al}{2}), and the lines are so strong
that the absorbing gas is very likely neutral (H $\approx$
\ion{H}{1}).  QSO-DLAs provide detailed information on the neutral
interstellar medium (ISM) of high redshift galaxies (Prochaska 2004
and references therein).

Our paper is structured as follows: in \S 2, we describe the
measurement of the heavy-element column densities in the \grb\
circumburst medium; in \S3 we derive the heavy-element depletion
pattern; in \S4 the visual extinction is estimated; in \S5 we
constrain, given the dust depletion, the extinction curve; and in \S6 we
summarize the results.

\begin{table}
\caption[t1]{Parameters for the absorption system at
$z=1.255$}\label{t1}
\begin{center} 
\begin{tabular}{lccc} 
\tableline\tableline&&&\\[-5pt] 
   &  $\log N$ & $b\ ^a$ & \\
Ion         &  [cm$^{-2}$] & (\kms)  &  [X/Zn] \\ 
[3pt]\tableline&&&\\[-7pt] 
\ion{Zn}{2}~~~~ & $13.54\pm0.06$ & 62\tablenotemark{b} & $\cdot\cdot\cdot$ \\
\ion{Si}{2} & $16.29\pm0.04$ &  $56\pm3$ & $-0.15\pm0.07$\\
\ion{Si}{2}$^*$ & $14.32\pm0.11$ &  56\tablenotemark{c} & $\cdot\cdot\cdot$\\
\ion{Mn}{2} & $13.62\pm0.03$ &  $81\pm27$ & $-0.66\pm0.07$\\
\ion{Cr}{2} & $13.95\pm0.03$ & 62\tablenotemark{b} & $-0.62\pm0.07$ \\ 
\ion{Fe}{2} & $15.48\pm0.04$ &  $62\pm1$ & $-0.91\pm0.07$ \\
\ion{Ni}{2} & $14.20\pm0.03$ & $43\pm5$ & $-0.94\pm0.07$ \\ 
\ion{Ti}{2} & $<12.9$\tablenotemark{d} & $\cdot\cdot\cdot$ & $<-1.0$ \\ 
\ion{Mg}{2} & $\sim15.8$  & 62\tablenotemark{b} &  $\sim-0.7$ \\
\ion{Mg}{1} & $13.26\pm0.01$ & $110\pm6$ &  $\cdot\cdot\cdot$ \\
\ion{Al}{2} & $\sim14.9$\tablenotemark{e} & 62\tablenotemark{b} & $\sim-0.4$ \\
\ion{C}{1} & $<13.7$\tablenotemark{d}  & $\cdot\cdot\cdot$ & $\cdot\cdot\cdot$ \\
\ion{Ca}{2} & $13.72\pm0.04$ & $\cdot\cdot\cdot$\tablenotemark{f} &  $\cdot\cdot\cdot$ \\
[2pt]\tableline
\end{tabular}
\tablenotetext{a}{Effective Doppler parameter.}
\tablenotetext{b}{Doppler parameter is assumed to be the same as for \ion{Fe}{2}.}
\tablenotetext{c}{Doppler parameter is assumed to be the same as for \ion{Si}{2}.}
\tablenotetext{d}{4 $\sigma$ upper limit, obtained assuming $b=62$ \kms.}
\tablenotetext{e}{Column density estimated using the curve of growth.}
\tablenotetext{f}{\ion{Ca}{2} shows three absorbing components.}
\tablecomments{Oscillator strengths are from: Bergeson \& Lawler (1993) for \ion{Zn}{2} and \ion{Cr}{2}; Verner et al. (1994) for \ion{Si}{2}; Schectman et al. (1998) for \ion{Si}{2}$^*$; Morton (1991) for \ion{Fe}{2}, \ion{Ti}{2}, \ion{Al}{2}, and \ion{Mg}{2}; Kling \& Griesmann 
(2000) for \ion{Mn}{2}; Fedchak et al. (2000) for \ion{Ni}{2}.}
\end{center}
\end{table}

\begin{figure}\label{f2}
\centerline{{\epsfxsize=7cm \epsfbox{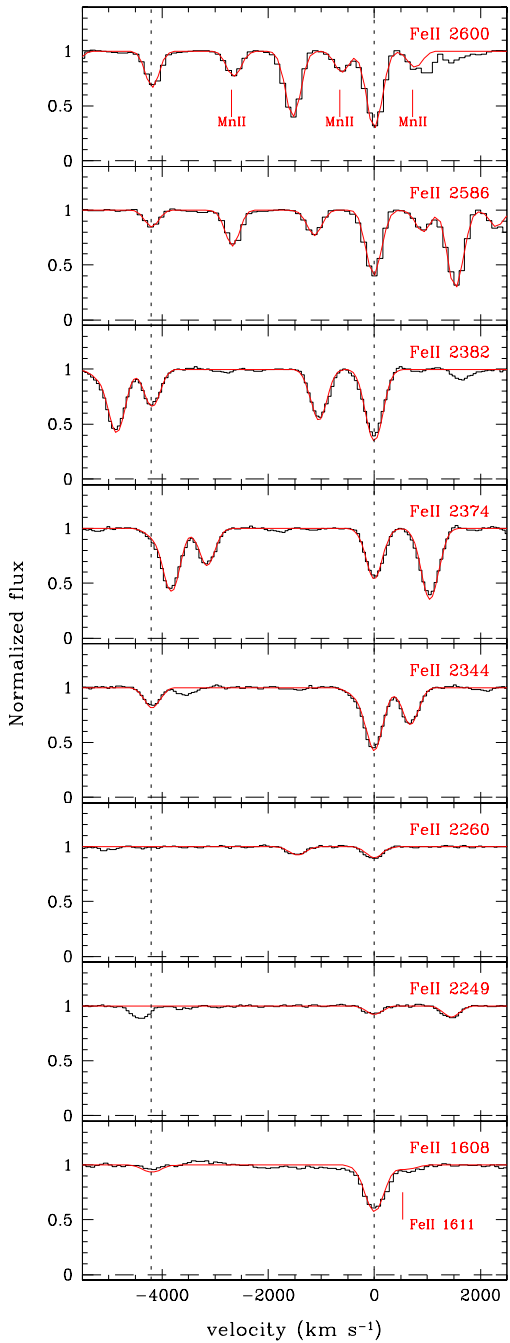}}}
\figcaption[f2]{Absorption lines of \ion{Fe}{2} in the GRB-DLA at
$z=1.255$ (at 0 \kms) and of the intervening system at
$z=1.224$ (at $-4200$ \kms) and their best-fit absorption profiles
({\it smooth lines}).  Nine \ion{Fe}{2} absorption lines are detected in the
GRB-DLA, allowing a good determination of the column density
(\feii~$=15.48\pm0.04$).  In the top panel, we also show the
best-fit \ion{Mn}{2} triplet absorption (\mnii~$=13.62\pm0.03$).}
\end{figure}

\begin{figure}\label{f3}
\centerline{{\epsfxsize=8.3cm \epsfbox{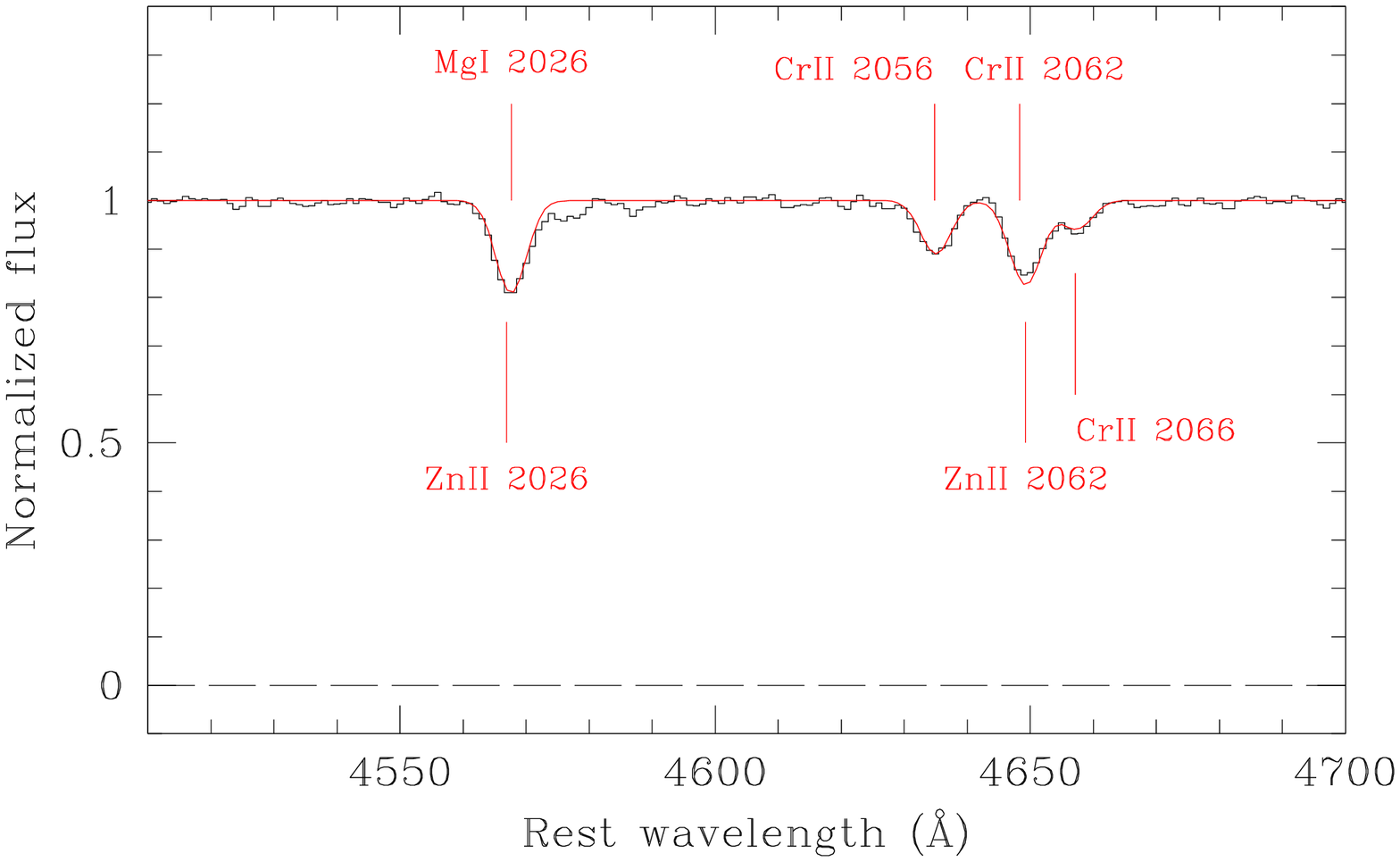}}}
\figcaption[f3]{Absorption lines of \ion{Zn}{2} and \ion{Cr}{2}.  The
smooth line is the best-fit absorption profile with $\log N_{\rm ZnII}
= 13.53$ and $\log N_{\rm CrII} = 13.95$. The Doppler parameter
is assumed to be the same as in \ion{Fe}{2} ($b=62$ \kms). If
the Doppler parameter is changed by $\pm15$ \kms, the best-fit
\ion{Cr}{2} and \ion{Zn}{2} column densities change by $\pm 0.01$
and $\pm0.02$ dex, respectively.  The contamination  of
\ion{Zn}{2} $\lambda2026$ by \ion{Mg}{1} $\lambda2026$ is taken into
account, and has been derived using the \ion{Mg}{1} $\lambda2852$ line
fit ($\log N_{\rm MgI} = 13.26$).  The \ion{Cr}{2} $\lambda2026$
absorption (not marked) is negligible.}
\end{figure}

\section{Measurements of Heavy-element Column Densities}

For the analysis of the GRB~020813 afterglow, we use the spectrum
obtained with the LRIS spectrograph on the Keck I telescope in 2002
August 13 (Barth et al.\ 2003).  The observations were made 4.65 h
after the {\it HETE-2} detection (Villase\~nor et al. 2002) when the visual
magnitude was $\sim19$. The observed spectral range is
$\lambda=3220-9150$ \AA, and the resolution is $\sim 1$ \AA~per pixel,
or FWHM $=94 - 32$ \kms\ from the blue to the red end.  The
signal-to-noise ratio in the $\sim 3$ hour total exposure is very
good, S/N $=50-100$, making this optical spectrum one of the best in
terms of resolution and S/N ever obtained for a GRB afterglow.  The
spectrum of \grb, after correction for Galactic extinction ($E_{B-V} =
0.111$; Schlegel et al. 1998), is shown in Figure~1.
Barth et al.\ (2003) reported the equivalent widths of detected
absorption lines, and identified two absorption systems, at $z=1.255$
and $z=1.224$ (see their Table 2). The former is very likely the
redshift of the GRB. The lower-redshift system intersects the GRB line
of sight, at a comoving distance\footnote{We adopt cosmological
parameters $h\equiv H_o/100= 0.7$, $\Omega_M = 0.3$, and
$\Omega_\Lambda = 0.7$.} of about 13 $h^{-1}$ Mpc from the GRB. We
fitted Voigt profiles to the absorption lines to determine column
densities for both absorption systems, and cross-checked the results
by using the curve-of-growth analysis (COG; Spitzer 1978) or the
apparent optical depth method (Savage \& Sembach 1991). As the
intervening system at $z=1.224$ has low metal column densities
($\sim60$ times lower than the $z=1.255$ system), we neglect its dust
extinction ($\lsim 0.01$ mag in the optical).

The absorption lines associated with the circumburst medium of \grb\
are numerous, and their equivalent widths $W_r$ in many cases deviate
little from the linear part of the COG (rest-frame $W_r<0.4$ \AA),
allowing a good determination of column densities. The main
uncertainty is the total number of absorbing components, which is not
known because of the finite resolution and  the intrinsic blending of
lines. Although the multiple absorbing components appear as a single
line, the single-component approximation still gives robust results
(Jenkins 1986; Savage \& Sembach 1991). However, in this case, the
width of the lines is an effective Doppler parameter
 and does not provide useful information on the temperature of
the gas.  Parameters of all lines relevant for this work
(and 1 $\sigma$ errors) and abundances relative to zinc are given in
Table 1.  We adopt the common assumption, valid for neutral gas
($N_{\rm H} \approx N_{\rm HI}$), that the abundances of ions with
ionization potentials above 13.6 eV, are very close to the total
abundance of the corresponding element: $N_{\rm Zn} \approx N_{\rm
ZnII}$, $N_{\rm Fe} \approx N_{\rm FeII}$, $N_{\rm Cr} \approx N_{\rm
CrII}$, etc. We discuss possible contamination  by
high-ionization species (\ion{C}{4} and \ion{Al}{3}) and ionization
corrections in \S\ref{ion}.

\subsection{Low-ionization Absorption Lines}

Figure~2 shows, in velocity space, the best-fit Voigt profiles for the
\ion{Fe}{2} absorption in the GRB-DLA and in the intervening system at
$z=1.224$, which marginally contaminates the GRB-DLA \ion{Fe}{2}
absorption. In the top and bottom panels, we also indicate the GRB-DLA
\ion{Mn}{2} triplet and the very weak \ion{Fe}{2} $\lambda1611$ line,
respectively. Since many \ion{Fe}{2} lines have large equivalent
widths, we estimate the uncertainty due to saturation by considering
the weakest transitions, \ion{Fe}{2} $\lambda\lambda2249,2260$ ($W_r
<0.3$ \AA).  Using the COG, we derive $\log N_{\rm FeII} =
15.5\pm0.1$ and $b>35$ \kms.  This column density is consistent with
(although less certain than) the one derived by profile fitting ($\log
N_{\rm FeII} = 15.48\pm0.04$ and $b=62\pm1$). Using the apparent
optical depth method, which gives reliable results when the line
profile is not known, we obtain $\log N_{\rm FeII}\simeq 15.47$, in
agreement with the profile-fitting result.

In Figure~3 we show the profile fits for \ion{Zn}{2}, \ion{Cr}{2}, and
\ion{Mg}{1}. The \ion{Mg}{1} $\lambda2026$ line marginally
contaminates \ion{Zn}{2} $\lambda2026$ and is constrained by
\ion{Mg}{1} $\lambda2852$ at $\lambda=6433.4$ \AA.  The Doppler
parameter $b$ for \ion{Zn}{2} and \ion{Cr}{2} is very uncertain and
thus has been assumed to be the same as for \ion{Fe}{2}. Changing $b$
by $\pm15$ \kms\ does not have a significant effect on the column
densities.  In general, the column densities of lines with small
equivalent widths depend only weakly on the Doppler parameter.
Figure~4 shows, in velocity space, the best-fit \ion{Si}{2}
profiles. Although the \ion{Si}{2} $\lambda1526$ line is contaminated
by the $z=1.224$ \ion{C}{4} $\lambda\lambda1548,1550$ doublet, the fit
is well constrained by the blue side of the line and by the
\ion{Si}{2} $\lambda1808$ absorption.

\begin{figure}\label{f4}
\centerline{{\epsfxsize=8.3cm \epsfbox{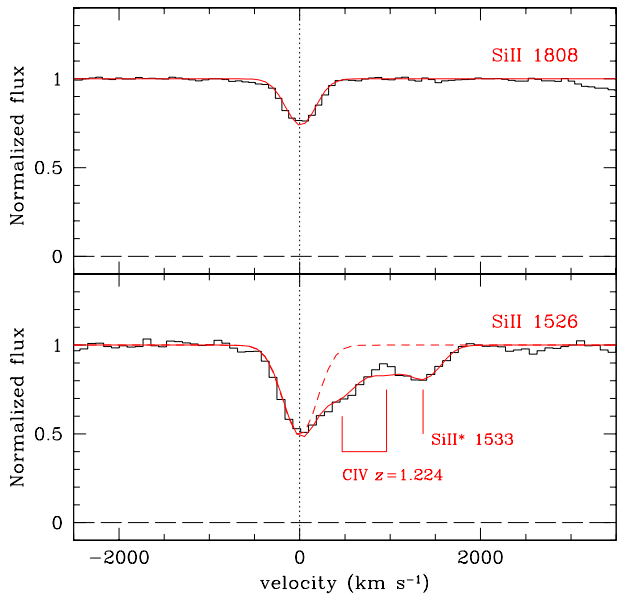}}}
\figcaption[f4]{Absorption lines of \ion{Si}{2} ($\log N_{\rm SiII} =
16.29$). The smooth line in the bottom panel also includes the best-fit
absorption profiles for the fine structure \ion{Si}{2}$^*$ $\lambda 1533$
absorption and the \ion{C}{4} doublet of the intervening system at
$z=1.224$.  We obtain a good fit to the \ion{Si}{2} column density
thanks to the isolated \ion{Si}{2} $\lambda 1808$ absorption and the
uncontaminated blue side of the \ion{Si}{2} $\lambda1526$ absorption
({\it dashed line}). }
\end{figure}

\begin{figure}\label{f5}
\centerline{{\epsfxsize=8.3cm \epsfbox{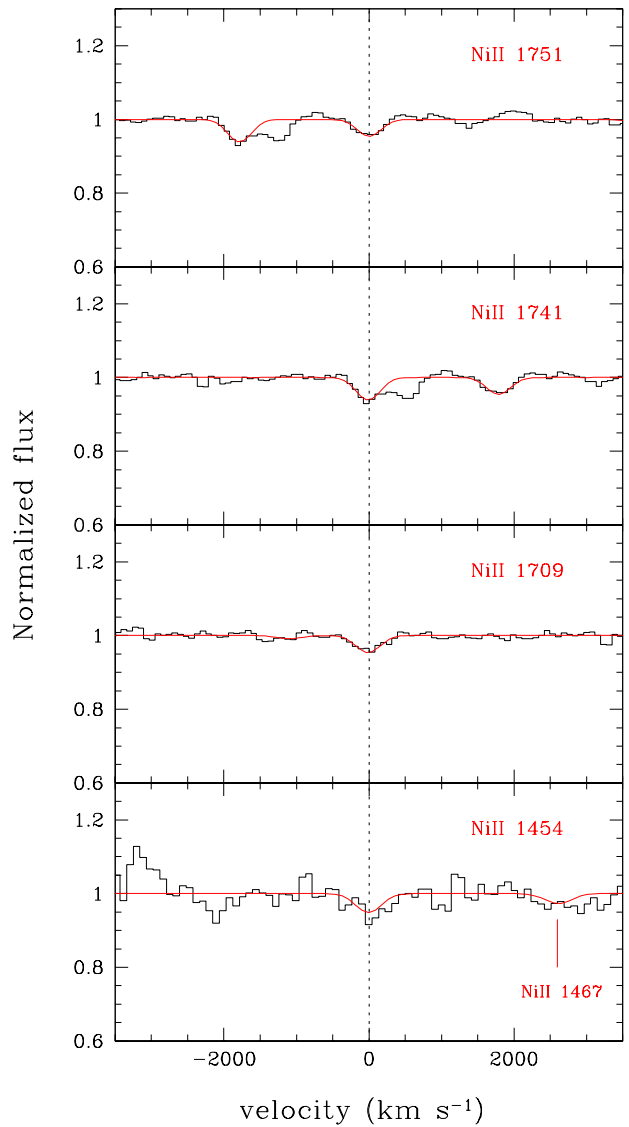}}}
\figcaption[f5]{Absorption lines of \ion{Ni}{2} ($\log N_{\rm NiII} =
14.20$). These weak lines have equivalent widths less than 0.2 \AA. If
the Doppler parameter is assumed to be the same as for \ion{Fe}{2}, the
\ion{Ni}{2} column density is 0.03 dex lower.}
\end{figure}

In Figure~5, we show 4 weak ($W_r<0.2$ \AA) \ion{Ni}{2} lines, not
identified by Barth et al.\ (2003), together with the best-fit
profile.  In the observed wavelength range, there are also three
\ion{Ti}{2} transitions. Two are not detected, while a third, the
stronger \ion{Ti}{2} $\lambda3384$ line, falls in the wavelength range
of the telluric absorption at $\sim7600$ \AA.  However, we can
still infer an interesting $4\sigma$ upper limit on the column
density.

The column densities of \ion{Mg}{2} and \ion{Al}{2} are very
uncertain because these lines are heavily saturated. The presence of
relatively strong \ion{Mg}{1} and \ion{Ca}{2} absorption (ionization
potential below 8 eV) is another indication that the gas is mainly
neutral. The non-detection of the relatively strong transition
\ion{C}{1} $\lambda 1656$ provides an upper limit for the column density.

\subsection{Higher Ionization Lines and the Ionization Correction}
\label{ion}

The spectrum of \grb\ shows the higher ionization lines of \ion{C}{4}
and \ion{Al}{3}. We  use these lines to demonstrate that the ionized
gas is a small fraction of the total and that the ionization
correction can be neglected.  \ion{Al}{3} and \ion{Al}{2} absorptions
were used to study the ionization state of the interstellar medium in
QSO-DLAs (Howk \& Sembach 1999).

The \ion{C}{4} column density is very likely a small fraction of the
total C column density in our GRB circumburst gas,
because the product of the wavelength and the oscillator strength
$\lambda f_\lambda$ for \ion{C}{4} $\lambda1551$ is twice that for
\ion{Fe}{2} $\lambda2374$, the cosmic abundance of C is 10 times higher
than that of Fe, and C is usually much less depleted than Fe in
dust ( $<10$ times). Therefore, if \ion{C}{4} gas were a large fraction
of the total C gas, we would expect the equivalent width of
\ion{C}{4} $\lambda1550$ to be much higher than that of
\ion{Fe}{2} $\lambda2374$. However, the observed equivalent widths are
1.31 and 1.42 \AA\ for \ion{C}{4} $\lambda1548$ and
\ion{Fe}{2} $\lambda2374$, respectively (Barth et al.\ 2003), showing
that the dominant ionization state of C is not \ion{C}{4}. Indeed we
estimate that the \ion{C}{4} column density is $\log N_{\rm C IV}
\simeq13.5$, compared with an expected total C column density (in gas
form) of $\log N_{\rm C} \approx 17$.

For Al, the ionization correction is less certain. The \ion{Al}{3}
doublet gives a column density of $\log N_{\rm Al III} \approx 14.0$
(if its $b$ value is the same as that of \ion{Fe}{2}). \ion{Al}{2}
shows only the very strong line at $\lambda=1670$ \AA, for which we
obtain $\log N_{\rm Al II} \approx 14.9$ (assuming $b$ as in
\ion{Fe}{2}), thus \ion{Al}{3} can only be a small correction of the
total Al.  If $b$ is left as a free parameter, the \ion{Al}{3} column
density is even smaller, although not by much. We also note once more
that most likely some \ion{Al}{3} is located in a different, more
ionized region in the host galaxy.

\begin{figure}\label{f6}
\centerline{{\epsfxsize=8.5cm \epsfbox{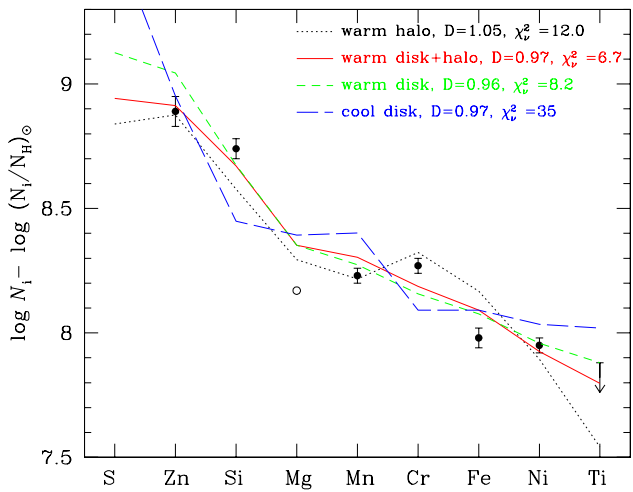}}}
\caption[f6]{Heavy-element column densities relative to solar
abundances in the \grb\ circumburst medium.  The dots are the measured
values. The broken lines are the best-fit expectations if the Galactic
dust depletion patterns are used as models. As a reference in the
models, S is not locked into dust grains in all patterns, while Zn is
largely depleted only in the cool disk clouds.  The upper limit on Ti
is shown but not used in the fit. The expected Ti values are from
Savage \& Sembach (1996) for the cool disk model, and are extrapolated
for the other depletion models assuming that Ti scales as Ni
(K. Sembach 2004, private communication). The measured Mg column
density is shown ({\it open circle}) but not used in the best fit,
because it is very uncertain. The dust-to-metals ratio, compared with
the Galactic value $D$, and the best-fit $\chi^2_{\nu}$ are also
reported.}
\end{figure}

\begin{table*}
\caption[t1]{Heavy-element column densities}\label{t2}
\begin{center} 
\begin{tabular}{lccccc} 
\tableline\tableline&&&&&\\[-5pt] 
&   & \multicolumn{4}{c}{$\log N^{tot}_i$ } \\
[4pt]\cline{3-6}\\[-4pt] 
{\sc Element} & $\log N^{obs}_i$ & WH\tablenotemark{a} & WDH\tablenotemark{b} & WD\tablenotemark{c} & CD\tablenotemark{d} \\
[3pt]\tableline&&&&&\\[-7pt] 
Zn & $13.54\pm0.06$ & $13.53\pm0.06$ & $13.56\pm0.06$ & $13.72\pm0.16$ & $14.23\pm0.09$ \\
Si & $16.29\pm0.04$ & $16.43\pm0.17$ & $16.46\pm0.08$ & $16.62\pm0.08$ & $17.13\pm0.29$ \\
Mn & $13.62\pm0.03$ & $14.27\pm0.03$ & $14.30\pm0.08$ & $14.46\pm0.05$ & $14.97\pm0.17$ \\
Cr & $13.95\pm0.03$ & $14.56\pm0.06$ & $14.59\pm0.09$ & $14.75\pm0.12$ & $15.26\pm0.18$ \\
Fe & $15.48\pm0.04$ & $16.38\pm0.19$ & $16.41\pm0.12$ & $16.57\pm0.10$ & $17.08\pm0.12$ \\
Ni & $14.20\pm0.03$ & $15.13\pm0.06$ & $15.16\pm0.04$ & $15.32\pm0.03$ & $15.83\pm0.09$ \\
\tableline&&&&&\\[-5pt] 
\end{tabular}
\tablenotetext{a}{Warm halo depletion pattern.}
\tablenotetext{b}{Warm disk+halo depletion pattern.}
\tablenotetext{c}{Warm disk depletion pattern.}
\tablenotetext{d}{Cool disk depletion pattern.}
\end{center}
\end{table*}

\subsection{The \ion{Si}{2}$^*$ Transition}

The spectrum of \grb\ also shows the ground fine-structure
\ion{Si}{2}$^*$ $\lambda1533$ transition, from which we derive

\begin{equation}\label{eq1b}
\frac{N_{\rm SiII^*}}{N_{\rm SiII}} = 10^{-1.97\pm0.12}\ .
\end{equation}

\noindent
The \ion{Si}{2}$^*$ absorption was detected by Barth et al.\ (2003),
but not identified.  The $N_{\rm SiII^*}/N_{\rm SiII}$ ratio can in
principle be used to determine the physical conditions of the gas
(Bahcall \& Wolf 1968).  Theoretical predictions as a function
of the gas temperature, gas density, and electron density have
been made and compared with observations in QSO spectra (Srianand \&
Petitjean 2000; Srianand \& Petitjean 2001; Silva \& Viegas 2002). The
$N_{\rm SiII^*}/N_{\rm SiII}$ ratio we find in this GRB-DLA is
$\sim100$ times lower than the value measured in a QSO highly ionized
region (Srianand \& Petitjean 2001). Although it is not possible to
derive the density of the gas without knowing its temperature, the
difference between the high-ionization gas in this QSO and \grb\
suggests a different physical nature.

The \ion{Si}{2}$^*$ transition was detected for the first time in a
GRB afterglow by Vreeswijk et al.\ (2004), who reported a low value of
the column density ratio, $N_{\rm SiII}/N_{\rm SiII^*}\approx
10^{-1.7}$. This GRB-DLA shows a very high \ion{H}{1} column density
($\log N_{\rm HI}=21.9$). Its gas is probably neutral, not only for
the large \ion{H}{1} column density, but also because the \ion{Al}{3} column
density (estimated by us using the reported doublet equivalent widths)
is more than 10 times smaller than the expected total column of Al. We
conclude that a low value of  the $N_{\rm SiII^*}/{N_{\rm SiII}}$
ratio (also found in our GRB-DLA) does not imply that the gas is
ionized. The presence of \ion{Si}{2}$^*$ is an important result that
deserves further attention to understand better the physical
conditions of the circumburst medium.

\section{Dust-Depletion Pattern}

The absorption lines detected in the GRB spectrum give the column
densities of heavy elements in the gas phase only.  Thus, as different
elements are locked in dust grains by different amounts, the column
densities measured from absorption lines carry the signature of the
chemical composition of the dust, i.e.\ the depletion pattern.

We have accurate column density measurements of six heavy elements:
zinc, chromium, silicon, manganese, iron and nickel, plus a stringent
upper limit for titanium (Table 1). To derive the dust-depletion
pattern, we compare these column densities with several dust-depletion
patterns in the Milky Way (Savage \& Sembach 1996): warm halo (WH),
warm disk+halo (WDH), warm disk (WD), and cool disk (CD).

We consider each of the four Milky Way (MW) depletion patterns as a model and
denote by $N_i^{exp}$ the expected column density of element $i$ in
such a model to be compared with the corresponding observed column
density $N_i^{obs}$ in the GRB spectrum.  In our approach, $N_i^{exp}$
is also a function of the dust-to-metals ratio relative to the
Galactic values ($D = d/d_{G}$):

\begin{equation}\label{eq2}
N_i^{exp} = N_i^{tot}[1+D(x_i-1)]\ .
\end{equation}

\noindent 
Here $x_i$ is the Galactic fraction of element $i$ in the gas phase and
$N_i^{tot}$ is the total (gas and dust phase) column density of the
same element in the GRB-DLA. To express $N_i^{tot}$, we 
define $y_i$ as the relative heavy-element abundance

\begin{equation}\label{eq2b}
N_i^{tot} = y_i\ N^{met}\ ,
\end{equation}

\noindent
where $N^{met}$ is the total column density of metals in the GRB-DLA.
For convenience, we express $y_i$ in terms of solar abundances
(Gravesse \& Sauval 1998): $y_i = c (N_i/N_{\rm H})_\odot$, where $c$
is a constant.  To obtain the best depletion model, we minimize the
reduced $\chi^2$

\begin{equation}\label{eq3}
\chi^2_\nu(D,N^{met}) = \frac{1}{n-2} \sum_{i=1}^n \left[\frac{\log N_i^{obs} - \log N_i^{exp}(D,N^{met})}{\sigma (\log N_i^{obs})}\right]^2,
\end{equation} 

\noindent
where $n=6$ is the number of elements in the fit and $\sigma (\log
N_i^{obs})$ is the error on the observed column density. We have
defined the $\chi^2$ using logarithmic quantities, because the
observational errors have an approximate lognormal distribution.
Using the observed column densities, we find four solutions (one for
each of the four depletion patterns) with four different $D$ and
$N^{met}$ values. In Figure~6, we plot the column densities relative to
solar abundances (both on a log scale) for the observed elements,
together with the four best-fit models.  The WH, WDH, and WD depletion
patterns give very similar $\chi^2_\nu$, whereas the CD pattern gives
a worse fit. The results are similar if the errors on all the column
densities are assumed to be equal. In Table 2, we report the total
(gas+dust) column densities $N_i^{tot}$ expected if the different
depletion patterns in the MW are taken as models. The errors listed in
Table 2 are the combination of 1 $\sigma$ measurement errors and the
deviations  between the expectations given by the depletion
models and the measured values.  In Table 3 we present the overall
dust parameters.

\begin{table}
\caption[t1]{Dust parameters}\label{t3}
\begin{center} 
\begin{tabular}{lcccc} 
\tableline\tableline&&&&\\[-5pt] 
Parameters & WH\tablenotemark{a} & WDH\tablenotemark{b} & WD\tablenotemark{c} & CD\tablenotemark{d} \\
\tableline&&&&\\[-5pt] 
$D$ & 1.05 & 0.97 & 0.96 & 0.97 \\
$\log (N^{met}/N^{met}_{\odot,21})$ & $-0.12\pm0.06$ & $-0.09\pm0.06$ & $0.07\pm0.08$ & $0.58\pm0.09$ \\
$A_V$  & $0.40^{+0.06}_{-0.05}$ & $0.40^{+0.06}_{-0.05}$ & $0.56^{+0.11}_{-0.09}$ & $1.84^{+0.43}_{-0.35}$ \\
[3pt]\tableline
\end{tabular}
\tablenotetext{a}{Warm halo depletion pattern.}
\tablenotetext{b}{Warm disk+halo depletion pattern.}
\tablenotetext{c}{Warm disk depletion pattern.}
\tablenotetext{d}{Cool disk depletion pattern.}
\end{center}
\end{table}

\begin{figure}\label{f7}
\centerline{{\epsfxsize=8.1cm \epsfbox{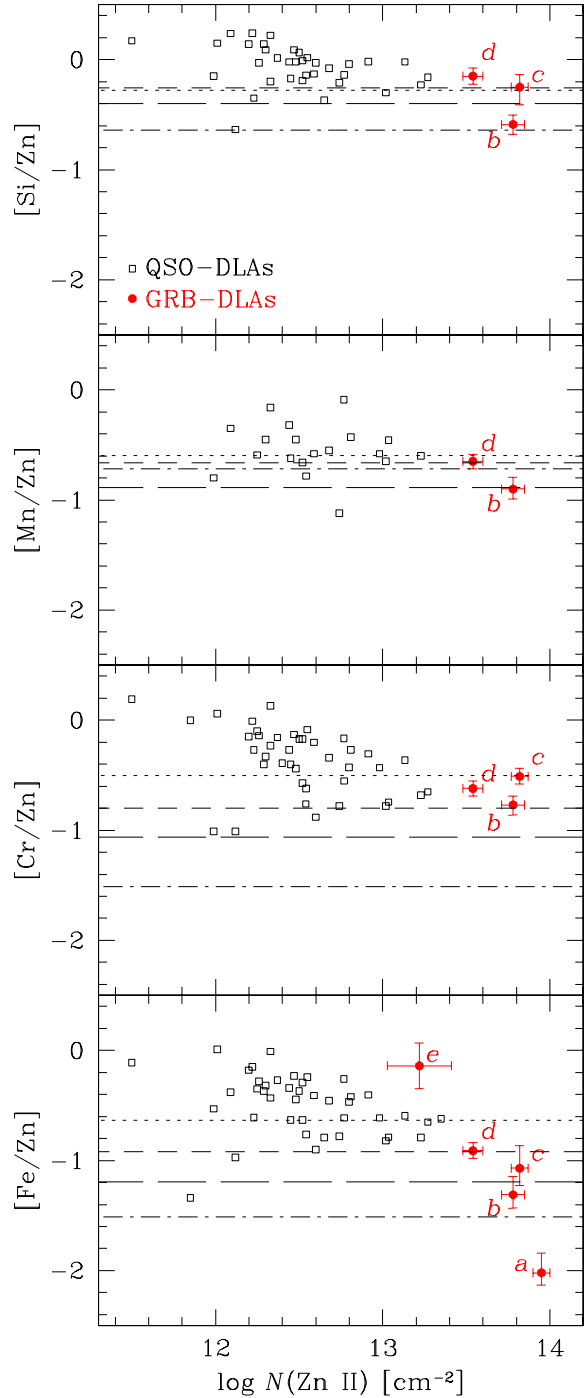}}}
\caption[f7]{Abundances of Fe, Cr, Mn, and Si relative to Zn,
vs. \ion{Zn}{2} column density in QSO-DLAs ({\it open squares}) and in
GRB-DLAs ({\it filled dots}). GRB-DLAs are: ($a$) GRB~990123, ($b$)
GRB~000926, ($c$) GRB~010222 (Savaglio et al.\ 2003), ($d$) GRB~020813
(this work), and ($e$) GRB~030323 (Vreeswijk et al. 2004). For
GRB~030323, the \ion{Zn}{2} column density is not measured, and we
used \ion{S}{2} instead, assuming [S/Zn] $\simeq 0$.  The mean errors
for [X/Zn] in QSO-DLAs are $\sim 0.08$ dex.  Dotted, short-dashed,
long-dashed, and dot-dashed horizontal lines are the warm halo, warm
disk+halo, warm disk, and cool disk values, respectively, reported by
Savage \& Sembach (1996).}
\end{figure}

\section{Extinction from Depletion Pattern}\label{opt-ext}

The extinction is proportional to the column of dust along the line of
sight.  In the solar neighborhood, a gas cloud with $10^{21}$ 
 \cm\ \ion{H}{1} column density has $A_V=0.5$
visual extinction (Bohlin et al. 1978). The value
of $A_V$ is about 1/4 lower in a Large Magellanic Cloud (LMC) cloud
with similar \ion{H}{1} column density (Prevot et al.\ 1984); however,
this is compensated by the lower metallicity (1/4 solar).  Therefore,
the visual extinctions for two clouds with similar columns of metals,
one in the LMC and another in the solar neighborhood, are similar. The
dust depletion in our GRB-DLA (where [Fe/Zn] $=-0.91\pm0.07$) is
closer to that in the LMC than that in the SMC ([Fe/Zn]
$=-1.12^{+0.37}_{-0.20}$ in the former, [Fe/Zn]
$=-0.57^{+0.09}_{-0.07}$ in the latter; Welty et al.\ 1997).

If the dust in the \grb\ circumburst medium were like that in the MW
or LMC, we would derive a visual extinction related to the column
density of dust $N^{dust}$ by

\begin{equation}\label{dust}
A_V = 0.5 \frac{N^{dust}}{N^{dust}_{\odot,21}} = 0.5 D \frac{N^{met}}{N^{met}_{\odot,21}}\ .
\end{equation}

\noindent
Here $N^{dust}_{\odot,21}$ and $N^{met}_{\odot,21}$ are the dust and
metal columns, respectively,  for a $N_{\rm HI}=10^{21}$ 
\cm\ gas cloud with solar metallicity, and $D$ and $N^{met}$, as
before, are the dust-to-metals ratio and the total metal column
in our GRB-DLA.  The $N^{met}/N^{met}_{\odot,21}$ and $A_V$ values
(and relative 1 $\sigma$ errors) are listed in Table 3. The inferred
$A_V$ values are all higher than 0.3 (95\%
confidence level), with a best fit, obtained for the WDH depletion
pattern (Fig.~6) of  $A_V=0.40^{+0.06}_{-0.05}$.

\begin{figure}\label{f8}
\centerline{{\epsfxsize=8.9cm \epsfbox{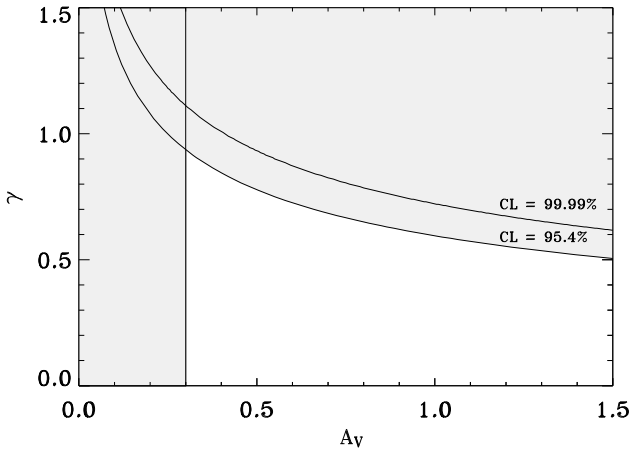}}}
\caption[f8]{Extinction curve index $\gamma$ and visual
extinction $A_V$ for constant confidence levels (CL), obtained by
comparing the reddened intrinsic GRB emission with the observed GRB
spectrum. The shaded vertical region indicates the $A_V$ values not
allowed by the depletion analysis ($95$\% CL).}
\end{figure}

\begin{figure}\label{f9}
\centerline{{\epsfxsize=8.5cm \epsfbox{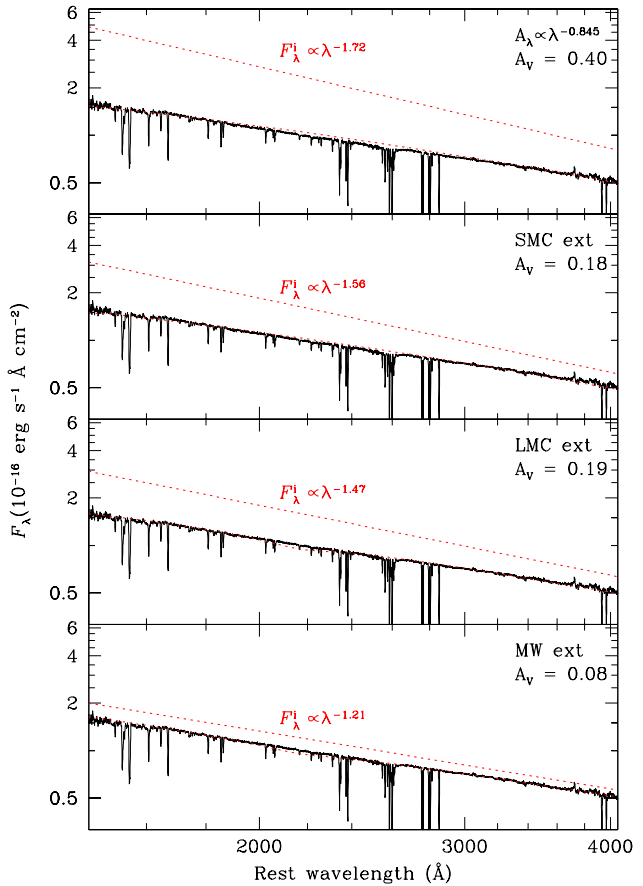}}}
\caption[f9]{Intrinsic and reddened  GRB spectrum ({\it upper and
lower dotted lines}) as compared with the observed spectrum ({\it solid
line}), for different extinction laws. In all cases, larger $A_V$ values
are not allowed ($95$\% CL). Note that the intrinsic GRB emission is
obtained by the best fit and is not assumed. In the top panel, the
extinction curve is a power law where  $\gamma=0.845$
is the maximum allowed value for  $A_V=0.40$.  
In the bottom three panels, MW, LMC, and SMC are assumed.}
\end{figure}

\subsection{QSO-DLAs and GRB-DLAs}

The QSO-DLAs are usually considered to be representative of the ISM of
the whole population of galaxies at high redshifts. The dust effects
in QSO-DLAs are generally small: the visual extinction in 44 QSO-DLAs
(with detected \ion{Zn}{2}, $0.39 < z < 3.39$) inferred using the same
approach as above, is always $A_V\lsim0.15$.  This difference from the
extinction in the \grb\ circumburst medium ($A_V>0.3$) is indicated by
the lower \ion{Zn}{2} column densities and the larger [Fe/Zn] values
typically seen in QSO-DLAs (zinc and iron are among the least and
most depleted elements, respectively). This is illustrated in the
bottom panel of Figure~7.

The difference between GRB-DLAs and QSO-DLAs may be the result of an
observational bias: QSO-DLAs are generally found in the foreground of
optically selected QSOs, and dusty QSO-DLAs are hard to find in such
samples (Fall \& Pei 1993). GRBs are (at least for some time) much
brighter than QSOs and are detected by gamma-rays and so might be much
less affected by this bias. Moreover, we know that GRBs are located in
star-forming regions, where the densities are presumably higher than
average, and large column densities of dust are not surprising.

The elements Cr, Mn, and Si (in addition to Fe) also indicate
relatively high dust depletion in \grb\ (Fig.~7). Similar high
depletions were found in the circumburst medium of three other
GRBs (Savaglio et al. 2003).  In another GRB-DLA (in
GRB~030323), the [Fe/S] is consistent with zero, indicating little or
no dust depletion (Vreeswijk et al. 2004). Since \ion{Zn}{2} is not
measured in this object, while \ion{S}{2} is, we assume [Zn/S] $\simeq
0$ and report this point in Figure~7.

\section{Extinction from Continuum Spectrum}

The presence of dust in the GRB circumburst medium can affect the
observed GRB continuum spectrum. The GRB extinction law has never been
determined by observation and is not known a priori. From the lack of
apparent reddening in many GRB spectra, it has been concluded
in those cases that dust is absent or that the extinction law is
a weak function of the wavelength (gray extinction; Stratta et al.\
2004). Perna et al. (2003) have shown, using
numerical simulations, that the GRB X-ray/UV radiation field can
significantly flatten the dust extinction curve.

We have already derived the heavy-element depletion pattern in \grb,
from which we inferred the visual extinction to be relatively high,
$A_V>0.3$, assuming the same rate of visual
extinction as in the MW or LMC per unit of metal column density.
However, the observed spectrum shows very little deviation from a
perfect power law and no sign of the 2200 \AA\
extinction feature (Fig.~1), which allows us to place strong
constraints on the extinction curve of the intervening dust.  We use a
fairly general approach, namely, we assume that the intrinsic GRB
emission is a power law

\begin{equation}\label{grb_i}
F_\lambda^i = F_V \left(\frac{5500}{\lambda}\right)^{\alpha}\ ,
\end{equation}

\noindent 
where $F_V$ and $\alpha$ are not fixed a priori, and we apply a dust
extinction law of the form

\begin{equation}\label{ext}
A_\lambda = A_V\left(\frac{5500}{\lambda}\right)^\gamma\ .
\end{equation}

\noindent
We vary $\alpha$, $F_V$, $\gamma$ and $A_V$, and compare the
corresponding model spectrum with the observed \grb\ spectrum using
$\chi^2$ minimization.  Because we are mainly interested in the
confidence levels (CL) of only two of these four parameters ($A_V$ and
$\gamma$), we project the four-dimensional confidence regions in the $A_V$
and $\gamma$ space.

In Figure~8 we show $\gamma$ versus $A_V$ for constant CL ({\it solid
lines}). For  $A_V=0.40$ (as inferred from the dust
depletion), any value of $\gamma$ between  $0.85$
and 0 is allowed (95\% CL), and this would give an intrinsic GRB
spectral slope $\alpha<1.72$. The shaded
vertical region is the one excluded by our depletion-pattern analysis.

We also calculated the maximum $A_V$ values (95\% CL) obtained with
MW, LMC, and SMC extinction curves. For the MW and LMC extinctions,
these give $A_V<0.08$ and $<0.19$, respectively, resulting from the
absence of the 2200 \AA\ extinction feature in the observed GRB
spectrum.  For the SMC extinction, which has a very weak 2200 \AA\
bump, the constraint is $A_V<0.18$.

In Figure~9, we show the intrinsic and reddened GRB model spectrum as
compared to the observed spectrum. For the MW and MC extinction
curves, this is done for the maximum allowed $A_V$ value  ({\it bottom
three panels}), while for the power-law extinction ({\it top panel}),
$A_V=0.40$ is assumed.  Since the upper limits on
the visual extinctions inferred from MW and MC reddening ($A_V<0.2$)
are smaller than the lower limits inferred from the depletions
($A_V>0.3$), we have reached a contradiction and are forced to
conclude that the dust in the circumburst medium of \grb\ cannot be
like that in the MW and MC in all respects (both chemical and optical
properties). We note that although variations in the
extinction law in different OB association stars have been found 
in the MW (Fitzpatrick \& Massa 1988), the presence of the 2200 \AA\
bump is still very clear and would not be consistent with the observed
GRB spectrum.

\section{Summary and Conclusions}

In this work, we studied, for the first time simultaneously, the
dust depletion pattern and the extinction law in the circumburst
medium of \grb. Very little is known about the dust properties
in the circumburst medium of any other GRB, mainly because the
emission flux declines so rapidly that optical spectra are very hard
to obtain.

An excellent optical spectrum of the \grb\ afterglow ($z=1.255$) was
obtained with the Keck I LRIS $\sim5$ hr after the burst by Barth et
al.\ (2003), when the GRB was still rather bright (19th mag in the
optical). From the numerous and strong absorption lines detected in
the spectrum, we measured the column density of six heavy elements
(Zn, Si, Cr, Mn, Fe, and Ni).  The relative abundances resemble the
Galactic dust-depletion patterns, from which we infer a visual
extinction $A_V\simeq 0.40$, and $A_V>0.3$ (95\% confidence).

The relatively high visual extinction in \grb\ is hard to reconcile
with the observed UV GRB continuum spectrum, which deviates very
little from a perfect power law, suggesting low reddening. However,
this can be explained if the extinction curve is only a weak function
of the wavelength.  We constrained the shape of the extinction curve,
assuming the power-law form $A_\lambda = A_V
(5500/\lambda)^\gamma$. For $A_V=0.40$, we obtain $\gamma<0.85$.  This
high visual extinction rules out the possibility of MW, LMC, or SMC
reddening, for which only $A_V<0.08,0.19, 0.18$, respectively, are
allowed. The MW and LMC extinctions are also constrained by the
absence of the 2200 \AA\ extinction feature.

A moderately high visual extinction is not surprising if GRBs
originate in regions of intense star formation and high density
(Hjorth et al. 2003a). However, under some conditions, the GRB itself
could destroy some of the dust grains (Fruchter et al. 2001).

Finally, heavy-element column densities in the circumburst medium of
this GRB are considerably higher than those typically observed in QSO
damped Ly$\alpha$ systems (about 10 times higher), which is likely the result
of a selection bias in the latter.  The comparison of the
dust-depleted element Fe, with the non-depleted element Zn, in the gas
phase, gives [Fe/Zn] $=-0.91\pm0.07$, whereas in 44 QSO-DLAs
($\langle z \rangle =2.0$) this is [Fe/Zn]~$=-0.53$ (0.28 dex
1 $\sigma$ dispersion), suggesting a higher dust content in the
circumburst medium of \grb.

\acknowledgments 

We are grateful to Aaron Barth and his collaborators for allowing us
to use their Keck spectrum of \grb.  We thank the anonymous referee,
Tim Heckman, Nicola Masetti, Simona Mei, Eliana Palazzi, Nino Panagia,
Patrick Petitjean, James Rhoads, and Ken Sembach for interesting
insights and Alessandra Aloisi for providing a table with the most
up-to-date values of oscillator strengths. S.~S. acknowledges generous
funding from the David and Lucille Packard Foundation.

{}

\end{document}